\documentclass[reprint,amsmath,amssymb,superscriptaddress,aps]{revtex4-2}
\usepackage{graphicx}
\usepackage{subfigure}
\usepackage{dcolumn}
\usepackage{tabularx}
\usepackage{bm}
\usepackage{braket}
\usepackage{gensymb}
\usepackage{color}

\begin{document}
	\preprint{APS/123-QED}
	\title{High-sensitivity Optical Microcavity Acoustic Sensor Covering Free Spectral Range}
	\author{Qi Song}
	\altaffiliation{These authors contributed equally to this work: Qi Song and Hongjing Li.}
	\affiliation{State Key Laboratory of Photonics and Communications, Institute for Quantum Sensing and Information Processing, School of Sensing Science and Engineering, Shanghai Jiao Tong University, Shanghai 200240, People's Republic of China}
	
	\author{Hongjing Li}
	\email{lhjnet2012@sjtu.edu.cn}
	\affiliation{State Key Laboratory of Photonics and Communications, Institute for Quantum Sensing and Information Processing, School of Sensing Science and Engineering, Shanghai Jiao Tong University, Shanghai 200240, People's Republic of China}
	\affiliation{Hefei National Laboratory, Hefei, 230088, People's Republic of China}
	\affiliation{Shanghai Research Center for Quantum Sciences, Shanghai, 201315, People's Republic of China}

	\author{Chengxi Yu}
	\affiliation{State Key Laboratory of Photonics and Communications, Institute for Quantum Sensing and Information Processing, School of Sensing Science and Engineering, Shanghai Jiao Tong University, Shanghai 200240, People's Republic of China}
	
	\author{Ding Wang}
	\affiliation{State Key Laboratory of Photonics and Communications, Institute for Quantum Sensing and Information Processing, School of Sensing Science and Engineering, Shanghai Jiao Tong University, Shanghai 200240, People's Republic of China}
	
	\author{Xingyu Wu}
	\affiliation{School of Artificial Intelligence, Beijing Normal University, Beijing, 100875, People's Republic of China}
	
	\author{Zhiqiang Liu}
	\affiliation{State Key Laboratory of Photonics and Communications, Institute for Quantum Sensing and Information Processing, School of Sensing Science and Engineering, Shanghai Jiao Tong University, Shanghai 200240, People's Republic of China}
	
	\author{Jingzheng Huang}
	\affiliation{State Key Laboratory of Photonics and Communications, Institute for Quantum Sensing and Information Processing, School of Sensing Science and Engineering, Shanghai Jiao Tong University, Shanghai 200240, People's Republic of China}
	\affiliation{Hefei National Laboratory, Hefei, 230088, People's Republic of China}
	\affiliation{Shanghai Research Center for Quantum Sciences, Shanghai, 201315, People's Republic of China}
	
	\author{Chuan Wang}
	\affiliation{School of Artificial Intelligence, Beijing Normal University, Beijing, 100875, People's Republic of China}
	
	\author{Guihua Zeng}
	\email{ghzeng@sjtu.edu.cn}
	\affiliation{State Key Laboratory of Photonics and Communications, Institute for Quantum Sensing and Information Processing, School of Sensing Science and Engineering, Shanghai Jiao Tong University, Shanghai 200240, People's Republic of China}
	\affiliation{Hefei National Laboratory, Hefei, 230088, People's Republic of China}
	\affiliation{Shanghai Research Center for Quantum Sciences, Shanghai, 201315, People's Republic of China}
	
	\begin{abstract}	
	Optical whispering gallery mode microcavity acoustic sensors have emerged significant potential in high-sensitivity acoustic signal detection, but the narrow dynamic range limits the application prospect. To address the challenge, we propose and experimentally demonstrate a high-sensitivity acoustic sensor in optical whispering gallery mode microcavity. The proposed sensor integrates an extended Mach-Zehnder polarization interferometer with postselection, extending the dynamic range to the free spectral range. The sensing regions can be categorized into phase-drastic and phase-enhanced regions, both of which yield an optimal acoustic response that surpasses traditional transmission method. Experiment results demonstrate improvements of 57.87 dB in detection sensitivity and 26 times in minimal detectable acoustic pressure over the transmission method. Moreover, the application of coherent state and heterodyne detection can enhance response amplitude, further improving the detection sensitivity. Given the wide application range for acoustic dispersion and dissipation response, as well as the performance advantages of high sensitivity and wide dynamic range, the proposed WGM sensor offers a promising solution for acoustic sensing. Potentially, the structural design may be adaptable to other application scenarios involving time-varying signals through optical microcavity sensing.
	\end{abstract}
	
	\maketitle
	
	\section{Introduction}
	Acoustic sensors detect external changes by analyzing the characteristics of acoustic or mechanical waves, including velocity, amplitude, frequency, and phase\cite{vellekoop1998acoustic,drafts2001acoustic}, which have been found significant applications in scientific research\cite{barish2018nobel,walter2020distributed,wu2017beat,chaplin2013asteroseismology,pelling2004local,zhang2021vibrational,tang2023single}, daily life\cite{wang2015ultra,ma2022wave,sun2023whispering,attia2019review,weber2016contrast}, and industrial production\cite{kotze2016performance, fischer2016optical,DWIVEDI20183690}, such as inherent vibrational spectra identification, photoacoustic imaging, non-destructive inspection, etc. Piezoelectric acoustic sensors, which convert acoustic or mechanical signals to electric signals via piezoelectric effect, are relatively mature and commonly used. However, the noise equivalent pressure (NEP) and spatial resolution are constrained by the sensor size, while the working bandwidth is restricted by the resonant frequency of the sensor material, resulting in a limited application prospect\cite{guggenheim2017ultrasensitive}. Alternatively, optical acoustic sensors, including optical interferometer\cite{wang2019sound,chang2020case,li2018heat} and optical microcavity\cite{vahala2003optical,cao2024ultrasound}, providing a promising solution by transforming mechanical or acoustic waves as optical signals for the advantages of enhanced sensitivity, anti-electromagnetic interference performance, and integration capability\cite{guggenheim2017ultrasensitive,cao2024ultrasound}. Especially, whispering Gallery Mode (WGM) microcavities with ultrahigh quality factor ($Q$ factor) and small mode volume, significantly enhance light-matter interactions and markedly improve sensing sensitivity\cite{foreman2015whispering,jiang2020whispering,cao2024ultrasound,li2021cavity}, which have garnered extensive attention, and demonstrated substantial potential in acoustic signal sensing applications\cite{anetsberger2009near,kim2017air,basiri2019precision,yang2020multiphysical,pan2020microbubble,shnaiderman2020submicrometre,westerveld2021sensitive,sun2022encapsulated,meng2022dissipative,tu2022underwater,yang2023micropascal,xing2023ultrahigh,tang2023single,sun2023whispering}.
	
	\begin{figure*}[htbp]
		\centering\includegraphics[width=0.7\textwidth]{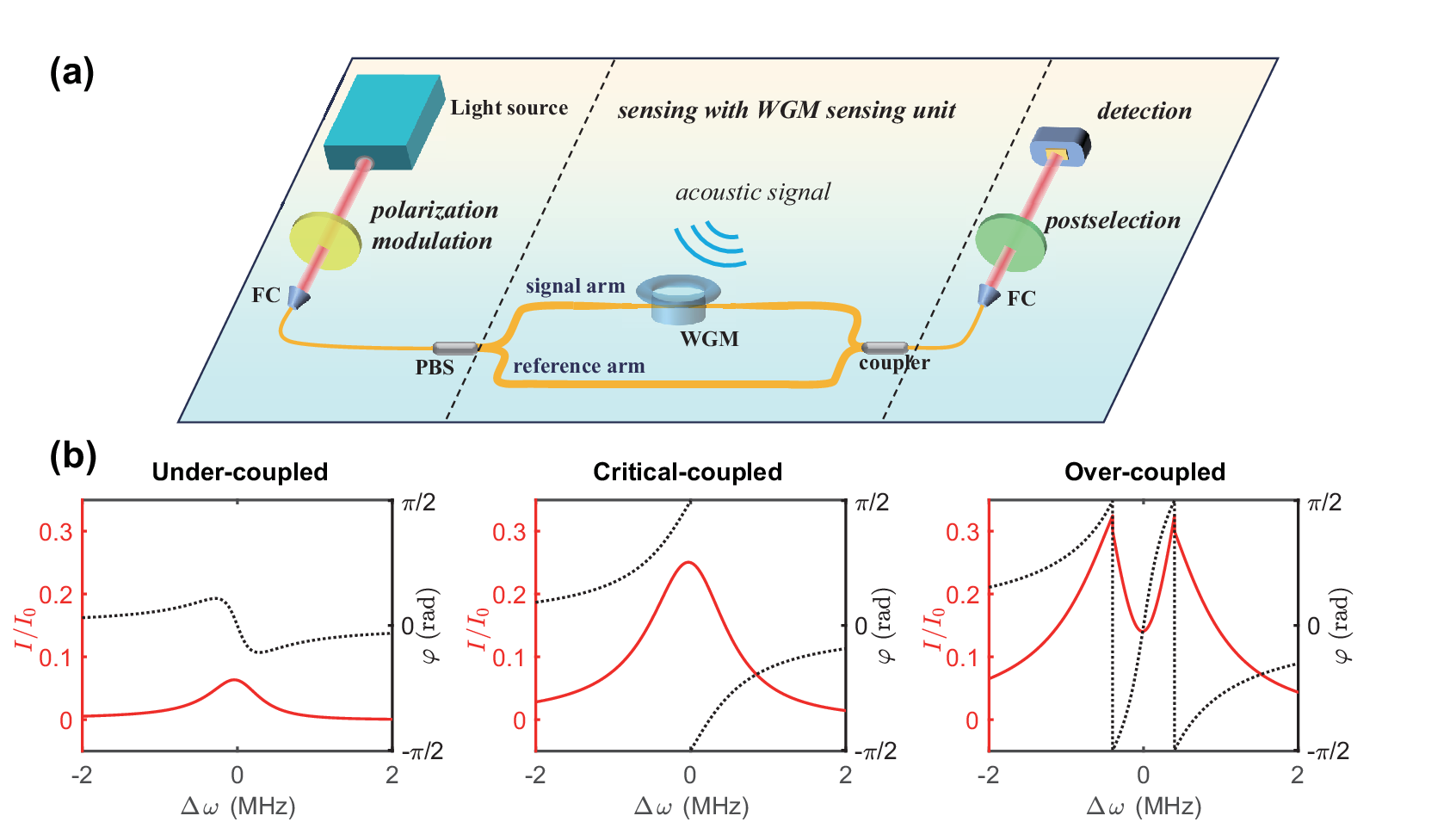}
		\caption{Conceptual design and sensing mechanism of the postselected WGM acoustic sensor. (a) Diagram of the postselected WGM acoustic sensor. FC is fiber coupler, and PBS is polarization beam splitter. (b) The relationship among $I/I_0$, $\varphi(t)$, and $\Delta \omega$ in under-coupled, critical-coupled, and over-coupled regimes, respectively. In simulation, the postselection angle is 0.05 rad, resonant wavelength is 1550 nm, and $Q$ factor is at the level of 10\textsuperscript{8}. Under-coupled regime: $\kappa_{0}=$0.6 MHz, $\kappa_{e}=$0.2 MHz; Critical-coupled regime: $\kappa_{0}=$0.6 MHz, $\kappa_{e}=$0.6 MHz; Over-coupled regime: $\kappa_{0}=$0.6 MHz, $\kappa_{e}=$1.0 MHz.}
		\label{fig1}
	\end{figure*}
	
	Previous researches on WGM acoustic sensors focus on achieving higher sensitivity by employing WGM microcavities with higher $Q$ factor, and acoustic signals can be retrieved by transmission method that tracks the transmission change of single cavity mode with probe light at a fixed wavelength around the resonant frequency\cite{foreman2015whispering,jiang2020whispering,cao2024ultrasound,li2021cavity}. The high $Q$ factor leads to a narrow bandwidth of at most twice full width at half maxima (FWHM), and enables highly sensitive detection of minute transmission changes, so that the dynamic range of the WGM microcavity is reduced by one-thousandth to ten-thousandth of the free spectral range (FSR) of different WGM microcavities with $Q$ factor of 10\textsuperscript{8}. In practical applications, unknown and stochastic acoustic signals may cause resonant shift exceeding the dynamic range, leading to information loss, and rendering the transmission method ineffective\cite{liao2021optical}. Multimode sensing method may provide a solution with larger dynamic range and less uncertainty, but applicability in acoustic signal sensing is restricted by the sweep period of the tunable laser, or the integration time of the spectrometer\cite{liao2021optical,duan2022high,lu2021experimental,zhang2021proposal}.
	
	In this paper, a high-sensitivity optical WGM microcavity acoustic sensor within FSR is proposed and experimentally demonstrated by utilizing an extended Mach-Zehnder polarization interferometer with postselection. Based on the relationship between phase characteristic and coupling strength, the sensing regions of the proposed approach can be divided into phase-drastic and phase-enhanced regions, and the optimal acoustic response exceeds that of traditional transmission method. In verification experiment, a 57.87 dB improvement in detection sensitivity, and a 26-fold enhancement in minimal detectable acoustic pressure (MDAP) over the transmission method is observed with the same devices. Additionally, the employment of coherent states and heterodyne detection can further enhance the response amplitude, potentially enabling the sensitivity up to the sub-micropascal level. The proposed WGM sensor provides a promising solution with high sensitivity and wide dynamic range for acoustic sensing, and the structure may also be applicable to other time-varying signal sensing via WGM microcavities, opening up various application scenarios.
	
	\section{Conceptual design and sensing mechanism}
	In order to realize the measurement of acoustic signal with wide dynamic range and high sensitivity based on optical microcavity, we propose a WGM acoustic sensor via extended polarization Mach-Zehnder interferometer with postselection, hereafter referred to as ``the postselected WGM acoustic sensor''. As depicted in Fig. \ref{fig1}(a), the sensing procedure can be divided into four steps, i.e., polarization modulation, sensing with WGM acoustic sensing unit, postselection and detection. In polarization modulation, the probe light is modulated as a linear polarization state of 45\degree. Different from previous researches, the WGM acoustic sensing unit is combined with a polarization Mach-Zehnder interferometer, where the coupled WGM microcavity is situated at the signal arm. Under the influence of the acoustic signal, the transmission and phase characteristics of the WGM microcavity change, resulting in difference between the signal arm and reference arm of the interferometer. In postselection, the evolved probe light is postselected by a preset polarization state, which is nearly orthogonal to 45\degree \space linear polarization state with a tiny postselection angle of $|\varepsilon|\ll1$. Given the real-time detection of acoustic signals, light intensity is detected to retrieve the signal. 
	
	\begin{figure*}[htbp]
		\centering\includegraphics[width=0.7\textwidth]{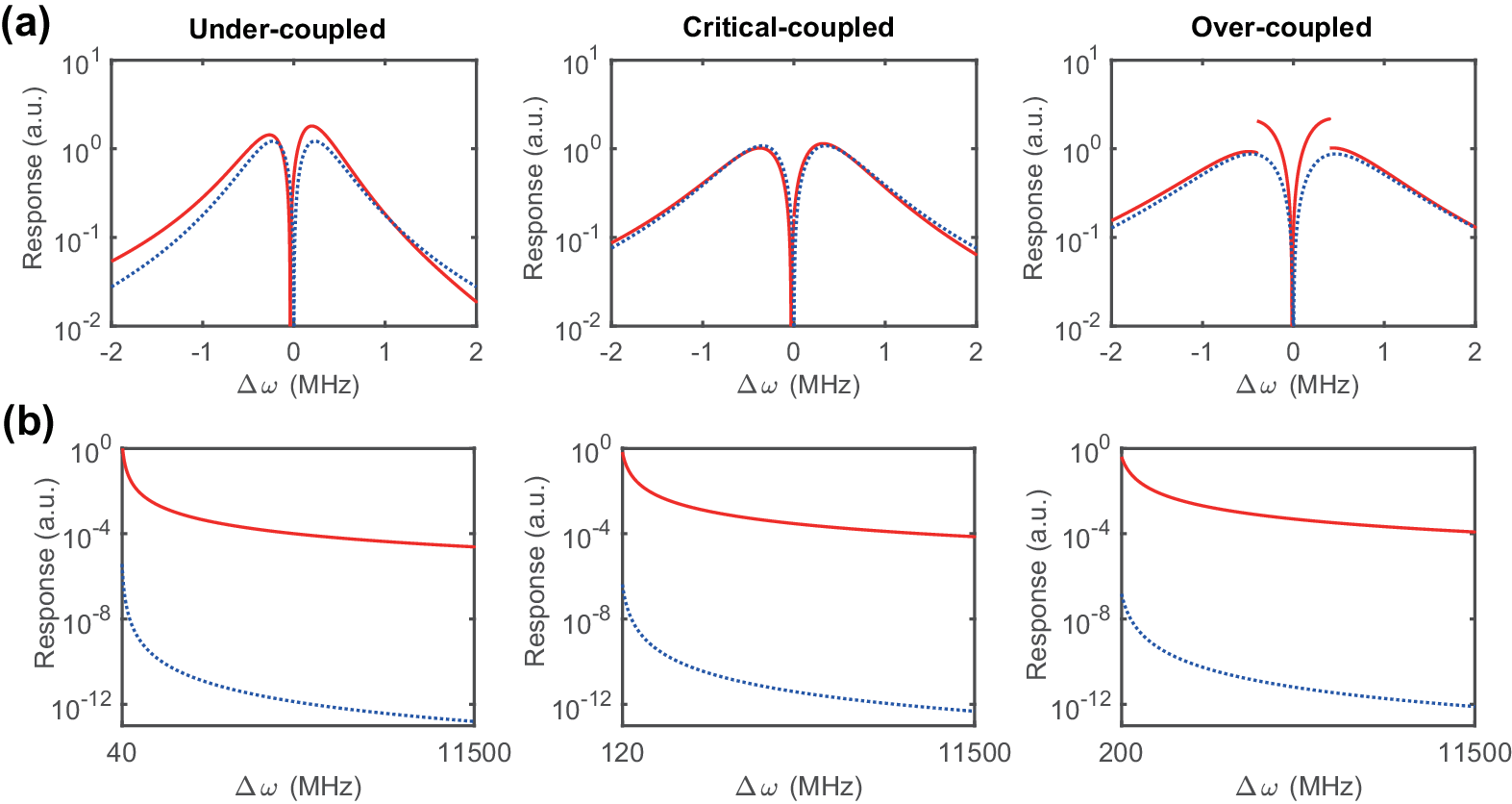}
		\caption{\textbf{Acoustic dispersive response of the postselected acoustic WGM sensor in different coupling regimes.} In simulation, the postselection angle is 0.05 rad, $Q$ factor is at the level of 10\textsuperscript{8}, FSR is 23 GHz, and resonant wavelength is 1550 nm. Under-coupled regime: $\kappa_{0}=$0.6 MHz, $\kappa_{e}=$0.2 MHz; Critical-coupled regime: $\kappa_{0}=$0.6 MHz, $\kappa_{e}=$0.6 MHz; Over-coupled regime: $\kappa_{0}=$0.6 MHz, $\kappa_{e}=$1.0 MHz. The red line and blue dashed line respectively represent the dispersive response of the postselected approach and transmission method. (a) Phase-drastic region; (b) Phase-enhanced region. }
		\label{fig2}
	\end{figure*}
	
	Consider that an acoustic wave $s(t)$ with amplitude $A$ and frequency $f_a$ interacts with the coupled WGM microcavity system, photoelastic effect and sound-induced strains cause the dispersive response, which leads to the shift of resonant frequency of the cavity mode\cite{meng2022dissipative,xing2023ultrahigh,balram2014moving,primo2020quasinormal}, i.e., $\omega_c(t)=\omega_0+ A_f A \sin (2\pi f_a t)$, where $\omega_0$ is the initial resonant frequency, and the resonant frequency shift due to unit acoustic pressure $A_f$ is determined by the particular microcavity and associated with the acoustic frequency $f_a$. In contrary, the vibration of tapered fiber induces the optical dissipative response, resulting in the change of the evanescent coupling between the modes of microcavity and tapered fiber\cite{meng2022dissipative,cao2024ultrasound}, i.e., $\kappa_{e}(t)=\kappa_{e}^0+ \delta_f A \sin (2 \pi f_a t)$, where $\kappa_{e}$ is the coupling dissipation of the cavity mode, $\kappa_{e}^0$ is the initial coupling dissipation of the cavity mode, $\delta_f$ is the change of the coupling dissipation under acoustic waves and associated with particular coupling system. Based on mode coupling theory\cite{gorodetsky1999optical}, the transmission characteristic and phase characteristic of cavity modes change, rendering the microcavity is sensitive to the acoustic signal. After postselection, the output light intensity is altered, i.e.,
	\begin{equation}
		\label{I}
		I(t)=\frac{I_0}{4}[1+T(t)-2\sqrt{T(t)}\cos(\varphi(t)+\varepsilon)],
	\end{equation}
	where both transmission characteristic $T(t)=[4\Delta\omega^2+(\kappa_{0} - \kappa_{e})^2]/[4\Delta\omega^2+(\kappa_{0} + \kappa_{e})^2]$ and phase characteristic $\varphi(t)=\arctan[(4\kappa_{e}\Delta\omega)/(\kappa_{e}^2-\kappa_{0}^2-4\Delta\omega^2)]$ of the cavity mode are utilized. Here, $I_0$ is the initial light intensity, $\kappa_{0}$ is the intrinsic loss, and detuning frequency $\Delta \omega$ is the difference of probe light frequency and resonant frequency. Subsequently, the acoustic signal can be retrieved when $A_f$, or $\delta_f $ is determined based on dispersive or dissipative sensing mechanisms. 
	
	The application of postselection leads to a complex scaling factor of $A_w$ on the observable system. When $|A_w(t)\varphi(t)/2 |\ll1$ is satisfied\cite{dressel2014colloquium,PhysRevA.76.044103}, Eq.(\ref{I}) can be approximated to 
	\begin{equation}
		\label{I2}
		I(t)\approx I_0 \xi [1+\varphi(t) \mathrm{Im} A_w],
	\end{equation}
	where $\xi=1/4[(\sqrt{T}-1)^2\cos^2(\varepsilon/2)+(\sqrt{T}+1)^2\sin^2(\varepsilon/2)]$ and 
	$\mathrm{Im} A_w=(2\sqrt{T}\sin\varepsilon)/[(\sqrt{T}-1)^2\cos^2(\varepsilon/2)+(\sqrt{T}+1)^2\sin^2(\varepsilon/2)]$, and $|\mathrm{Im} A_w|>1$ lead to an amplification factor on $\varphi(t)$. Based on the two conditions, the sensing regions of the postselected WGM acoustic sensor can be divided into phase-drastic and phase-enhanced regions (See details in the Appendix A). 
	
	As shown in Fig. \ref{fig1}(b), the phase-drastic region is typically located around the resonant frequency, and $I/I_0$ can be viewed as a function of detuning frequency $\Delta \omega$ in different coupling regimes. In under-coupled regime ($\kappa_e<\kappa_0$), continuous $\varphi(t)$ leads to continuous $I/I_0$ with a maximum value of $M\approx\kappa_e^2/(\kappa_e+\kappa_0)^2$ In critical-coupled regime ($\kappa_e=\kappa_0$), a discontinuity point at $\Delta \omega=0$ on $\varphi(t)$ results in a maximum value of $M=1/4$ on $I/I_0$. In the over-coupled regime ($\kappa_e>\kappa_0$), discontinuity points of $\Delta \omega=\pm\sqrt{\kappa_{e}^2-\kappa_{0}^2}/2$ emerge on $\varphi(t)$, and subsequently two inflection points on $I/I_0$ with a maximum value of $M\approx\kappa_e/[2(\kappa_e+\kappa_0)]$. 
	
	The phase-enhanced regions are symmetry with resonant frequency, which are situated far from the resonant frequency, and generally cover FSR. In this case, the two conditions of $|A_w(t)\varphi(t)/2 |\ll1$ and $|\mathrm{Im} A_w|>1$ are satisfied, and the output light intensity has a linear relationship with $\varphi(t)$, as expressed by Eq. (\ref{I2}). It indicates that $\varphi(t)$ can be extracted with an amplification factor of $\mathrm{Im} A_w$ above the noise proportional to light intensity, such as relative intensity noise, device imperfections, and thermal noise\cite{brunner2010measuring,song2023weak}. 
	
	\section{Response enhancement covering free spectral range}
	
	Both dispersive and dissipative responses can be utilized to retrieve the acoustic signal, acoustic response and dynamic range are chosen as figures of merit to evaluate the performance of the postselected WGM acoustic sensor, and the acoustic response can represent the detection sensitivity\cite{meng2022dissipative,xing2023ultrahigh}. Without loss of generality, we take acoustic dispersion response as an example for performance analysis, and one can refer to Appendix B for the performance of acoustic dissipative response. 
	
	The acoustic dispersion response of phase-drastic region can be defined as 
	\begin{equation}
		\label{r1}	
		R_1 = - A_f \frac{\partial (I/I_0)}{M \partial\Delta \omega (t)}\frac{\partial \Delta \omega (t) }{\partial s},
	\end{equation}
	where $\partial \Delta \omega (t)/ \partial s$ is the sensitivity coefficient of the microcavity to acoustic waves, and determined by a particular WGM microcavity. Further comparison with the transmission method is conducted, whose acoustic dispersion response can be expressed as $R_0 =\partial T/\partial \Delta \omega (t)\cdot\partial \Delta \omega (t)/\partial s$\cite{meng2022dissipative,cao2024ultrasound,xing2023ultrahigh}. As depicted in Fig. \ref{fig2}(a), the simulation results of cavity mode with a $Q$ factor of about 10\textsuperscript{8} have a stronger response than the transmission method, which could surpass one order of magnitude in over-coupled regime. 
	
\begin{figure}[hbpt]
		\centering\includegraphics[width=0.35\textwidth]{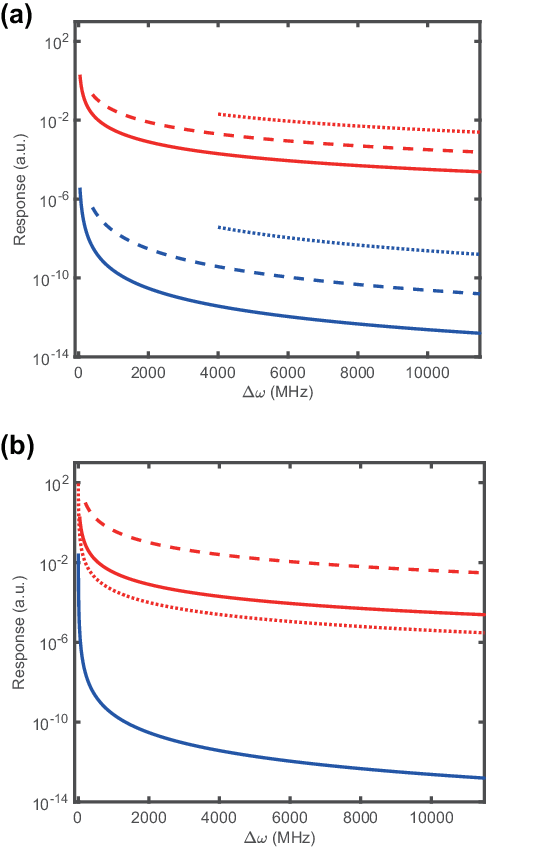}
		\caption{Acoustic dispersion response in under-coupled condition of under different $Q$ factors and postselection angles. In simulation, FSR is 23 GHz, and resonant wavelength is 1550 nm. (a) Acoustic dispersion response of different $Q$ factors at $\varepsilon$=0.05 rad. The red and blue lines show the results of the postselected sensor and transmission method, while the solid, dashed and dotted lines show the results of $Q$ factor at the level of $10^8$ ($\kappa_{0}=$0.6 MHz, $\kappa_{e}=$0.2 MHz), $10^7$ ($\kappa_{0}=$6 MHz, $\kappa_{e}=$ 2 MHz) and $10^6$($\kappa_{0}=$ 60 MHz, $\kappa_{e}=$ 20 MHz). (b) Acoustic dispersion response of different postselection angles when $Q$ factor is at the level of $10^8$. The dotted line, solid line, and dashed line in red respectively show the results of $\varepsilon$=0.1 rad, $\varepsilon$=0.05 rad and $\varepsilon$=0.01 rad, while the blue line shows the result of transmission method.}
		\label{fig3}
	\end{figure}

	Correspondingly in phase-enhanced regions, the acoustic response can be approximated to 
	\begin{equation}
		\label{r2}	
		R_2 = - A_f \frac{\mathrm{Im} A_w}{\xi}\frac{\partial \varphi(t)}{\partial\Delta \omega (t)}\frac{\partial \Delta \omega (t)}{\partial s}.
	\end{equation}
	The simulation results in Fig. \ref{fig2}(b) show the improvement could reach 7 orders of magnitude than that of the transmission method when the cavity mode with a $Q$ factor is about 10\textsuperscript{8}. It is worth to note that both of the setting of postselection angle and coupling strength determine the response and dynamic range of the phase-enhanced region, and the setting of postselction is further constrained by the coupling strength as restricted by the two conditions of $|A_w \varphi(t)/2|\ll1$ and $|\mathrm{Im} A_w| >1$. Without loss of generality, we take the acoustic dispersion response in the under-coupled condition as an example to investigate the effects. As shown in Fig. \ref{fig3}(a), a cavity mode with a higher $Q$ factor typically exhibits a smaller phase in regions far from the resonant peak. This characteristic enables a smaller postselection angle and results in a wider phase-enhanced region at a given postselection angle. 
	Conversely, a smaller postselection angle leads to a narrower phase-enhanced region but with a higher response, as depicted in Fig. \ref{fig3}(b). Therefore, the postselection angle should be determined based on coupling condition in practice to make it suitable for the requirements of actual experiments regarding sensitivity and dynamic range.
	
	By integrating the results in phase-drastic region and phase-enhanced regions, the postselected approach demonstrates a stronger optimal dispersion response than the transmission method, and the dynamic range can be extended to be close to FSR. For transmission method, the dynamic range is at most twice FWHM of the cavity mode. Correspondingly, the improvement factor on dynamic range than transmission method can be calculated by 
	\begin{equation}
		\label{dr}
		DR_{\mathrm{im}}=\frac{c/(n\pi D)}{2\times 2\pi c/(\lambda_c Q)}=\frac{\lambda_c Q}{4 \pi^2 n D },
	\end{equation} 
	where $c$ is the light velocity, $Q$, $n$, $D$, and $\lambda_c$ are the $Q$ factor, effective refractive index, diameter, and resonant wavelength of the WGM microcavity, respectively. According to Eq. (\ref{dr}), the improvement factor is in direct proportion to the $Q$ factor, which could reach 3 to 4 orders of magnitude for a cavity mode with a $Q$ factor of 10\textsuperscript{8}.

	\section{Experiment results and discussion}
\begin{figure*}[htbp]
		\centering\includegraphics[width=0.72\textwidth]{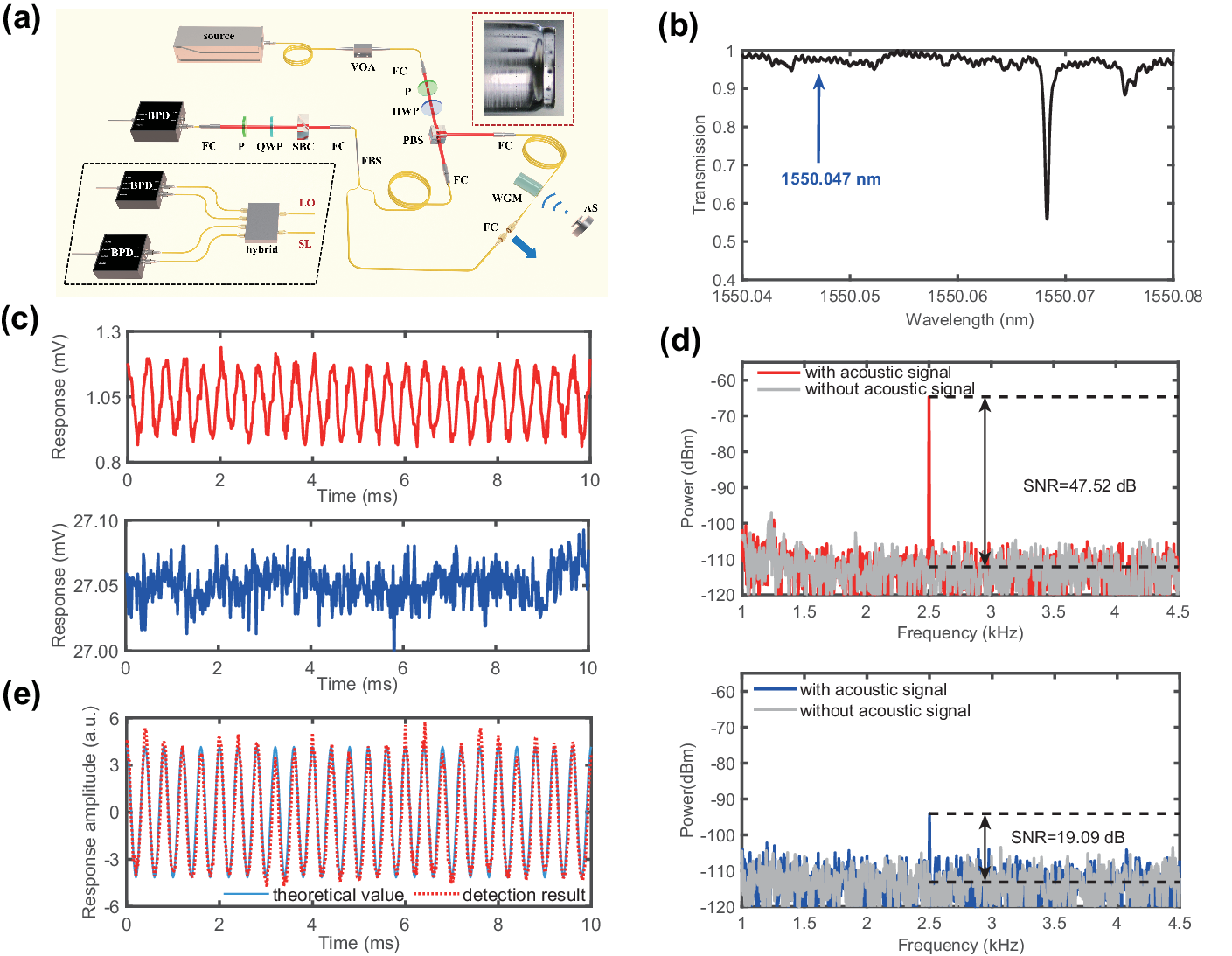}
		\caption{Experimental Setup and detection results.(a) Experimental setup. The MgF\textsubscript{2} microcavity is framed by a red box, and the abbreviations are as follows, VOA: variable optical attenuator; FC: fiber coupler; P: polarizer; HWP: half wave plate; PBS: polarization beam splitter; FBS: fiber beam splitter; SBC: Soleil-Babinet Compensator; QWP: quarter wave plate; BPD: balanced photoelectric detector; AS: acoustic signal source; WGM: the coupled MgF\textsubscript{2} microcavity; LO: local oscillator; SL: signal light. (b) Transmission spectrum. The $Q$ factor is about 3$\times$10\textsuperscript{6}. (c) Response amplitude of sinusoidal acoustic signal at 2.5 kHz. In experiment, the postselected WGM acoustic sensor and transmission method are performed at the same time, and the blue arrow in (a) indicates the position that senses acoustic signal with transmission method. The received light intensity of the postselected WGM acoustic sensor with a postselection angle of about 0.07 rad is about 14 dB weaker than that of transmission method. The red line shows the response of the postselected WGM acoustic sensor, and the blue line shows that of transmission method. (d) Power spectral density with and without acoustic signal. The red line shows the power spectral density of the postselected WGM acoustic sensor, and the blue line shows that of transmission method. (e) Response amplitude of the postselected WGM acoustic sensor using quantum coherent state. The implementation of heterodyne detection is framed by a black box in (a).}
		\label{fig4}
	\end{figure*}

	For experimental demonstration, we establish the postselected WGM acoustic sensor shown in Fig. \ref{fig4}(a). A polarization Mach-Zehnder interferometer incorporating a coupled MgF\textsubscript{2} microcavity serves as the acoustic sensing unit, where the MgF\textsubscript{2} microcavity is coupled with a tapered fiber with diameter of at the level of micrometer. Moreover, the diameter of the MgF\textsubscript{2} microcavity is about 3 mm, which is fabricated from a Z-cut MgF\textsubscript{2} crystal by turnings and chemical polishing. Firstly, polarization modulation is achieved using a polarizer with its optical axis oriented at 45\degree. Subsequently, after interaction with the WGM sensing unit, the probe light undergoes postselection via a quarter wave plate (QWP) with an optical axis at 45\degree and a polarizer with an optical axis at -47\degree. To simultaneously detect the acoustic signal using both the transmission method and the postselected WGM acoustic sensor, a 90:10 fiber beam splitter is inserted into the signal arm of the polarization Mach-Zehnder interferometer, indicated by a blue arrow. The output light from the 10\% path is directly detected using the transmission method, while the 90\% path is directed into the postselected approach. Additionally, a half wave plate (HWP) and a Soleil-Babinet compensator (SBC) are inserted to compensate for intensity and phase differences between the two arms. 
	
	Specifically, we focus on the phase-enhanced region, which is far away from the resonant frequency. By adjusting the coupling strength of the tapered fiber and the MgF\textsubscript{2} microcavity, transmission spectrum shown in Fig. \ref{fig4}(b) is generated, and probe light at a wavelength of 1550.047 nm is adopted to perform the verification experiment. For acoustic source, sinusoidal signals with different frequencies are applied, which are generated by a resonant horn driven by arbitrary waveform generator (AWG) at a distance of approximately 60 cm away from the coupled microcavity, and calibrated by a hydrophone. As depicted in Fig. \ref{fig4}(c), the response amplitude of postselected WGM acoustic sensor to sinusoidal acoustic signal is clearly discernible, whereas that of the transmission method is nearly indistinguishable from the background noise. Although the received light intensity of the postselected WGM acoustic sensor is 14 dB weaker than transmission method, it exhibits a significantly superior response amplitude, achieving an improvement of about 30 times over the transmission method. To quantitatively evaluate the response, the detection sensitivity can be calculated by $S_{\mathrm{im}}=20\mathrm{lg}(A^p/A^p_T)+20\mathrm{lg}(I/I_T) (\mathrm{dB})$, as the detection sensitivity is directly proportional to the received light intensity, where $A^p$ is the response amplitude, $I$ is the received light intensity, and subscript $T$ denotes the transmission method. Thus, an improvement of 57.87 dB over the transmission method in detection sensitivity is obtained. In practical coupling scenario, additional cavity mode may emerge, and the $Q$ factor of cavity modes influences the performance of the postselected WGM acoustic sensor, but it still achieves an improvement of nearly 3 orders of magnitude compared to the transmission method. MDAP is commonly chosen as a figure of merit to evaluate practical performance, which can be calculated by $P_{\mathrm{min}}(f_a)=1/\sqrt{\Delta f W_{\mathrm{SNR}}}P_{\mathrm{app}}(f_a)$\cite{yang2023micropascal,meng2022dissipative,xing2023ultrahigh}, where $P_{\mathrm{app}}(f_a)$ is the applied acoustic pressure at acoustic frequency of $f_a$, $\Delta f$ is the resolution bandwidth, and $W_{\mathrm{SNR}}$ is the ratio of power spectral density with and without signal. As shown in Fig. \ref{fig4}(d), a SNR of 47.52 dB with a resolution bandwidth of 5 Hz can be observed, corresponding to an MDAP value of 7.30 mPaHz\textsuperscript{-1/2}. The SNR is 28.43 dB higher than 19.09 dB of transmission method, resulting in an improvement over 26 times on MDAP.
	
	Indeed, the detectable acoustic pressure level is also determined by the minimum resolution of the detection equipment, manifesting as a few discrete values when variations are too subtle to be distinguished. To further enhance sensitivity, an additional experiment was conducted using coherent states and heterodyne detection. The coherent state was prepared by attenuating the signal light using a variable optical attenuator (VOA), with the intensity of the received signal light being approximately -70 dBm. Heterodyne detection was performed using local oscillator light, which is split from the same light source as the signal light, via a free-space passive optical hybrid equipped with two balanced photodetectors. As illustrated in Fig. \ref{fig4}(e), the detection results clearly distinguish the signal and closely match the theoretical values. Given that the received light intensity is approximately 30 dB weaker than that of classical probe light, the response amplitude demonstrates an improvement of 30.07 dB, potentially enhancing sensitivity to the sub-micropascal level.
	
	\begin{figure*}[hbpt]
		\centering\includegraphics[width=0.7\textwidth]{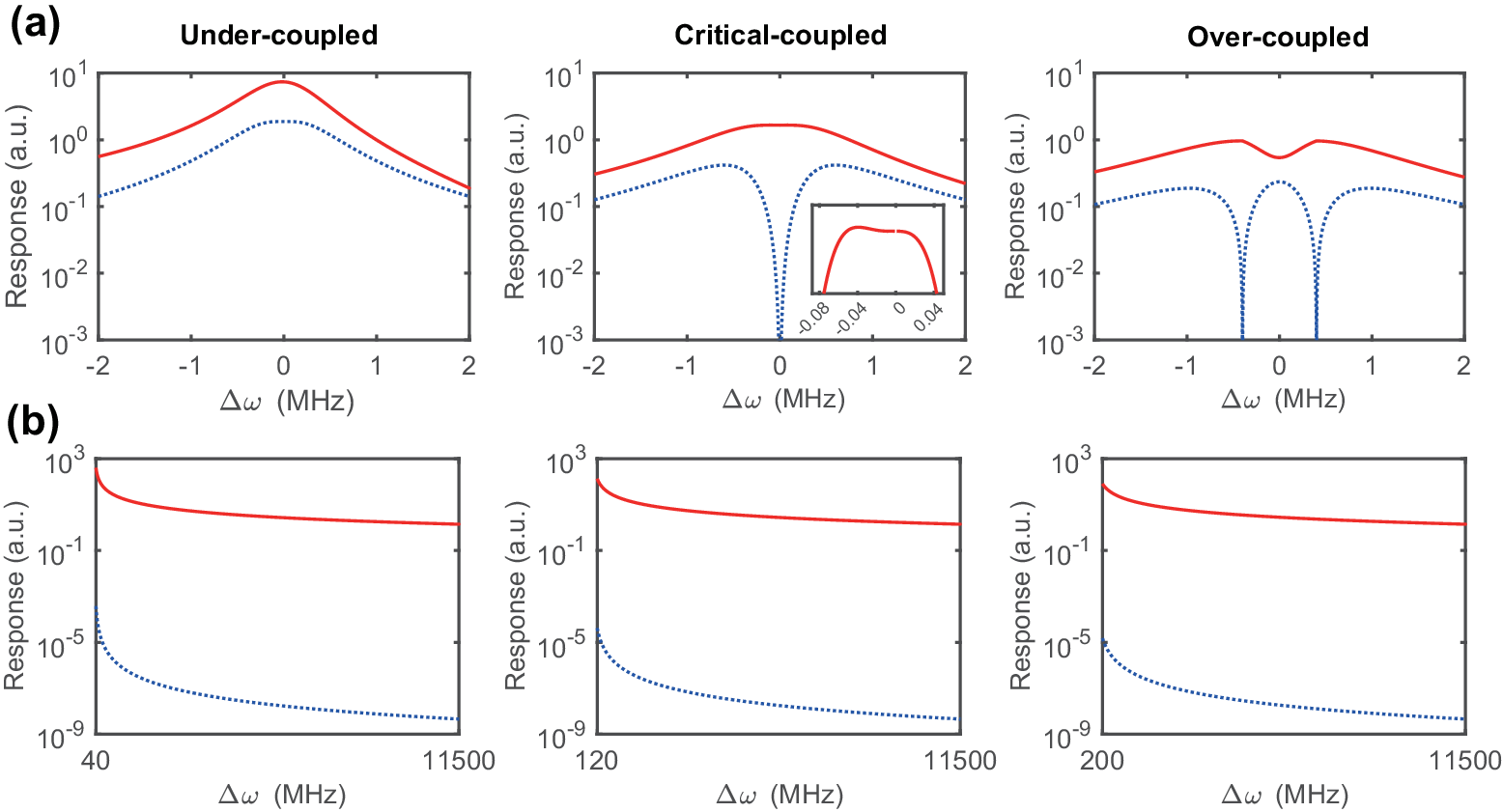}
		\caption{Acoustic dissipative response of the postselected WGM acoustic sensor in different coupling regime. In simulation, the postselection angle is 0.05 rad, $Q$ factor is at the level of 10\textsuperscript{8}, FSR is 23 GHz, and resonant wavelength is 1550 nm. Under-coupled regime: $\kappa_{0}=$0.6 MHz, $\kappa_{e}=$0.2 MHz; Critical-coupled regime: $\kappa_{0}=$0.6 MHz, $\kappa_{e}=$0.6 MHz; Over-coupled regime: $\kappa_{0}=$0.6 MHz, $\kappa_{e}=$1.0 MHz. The red line and blue dashed line respectively represent the dissipative response of the postselected approach and transmission method. (a) Phase-drastic region. (b) Phase-enhanced region.}
		\label{fig5}
	\end{figure*}
	
	\section{Conclusion}
	In summary, we have proposed and experimentally demonstrated a postselected WGM acoustic sensor to address the challenge that sensing acoustic signal with both high sensitivity and wide dynamic range. Different from the transmission method, the proposed WGM sensor incorporates an extended Mach-Zehnder polarization interferometer with postselection. The sensing regions can be categorized into phase-enhanced and phase-drastic regions, with an optimal response exceeding that of transmission method, while extending the dynamic range to reach FSR. In verification experiment, a 57.87 dB improvement in detection sensitivity over the transmission method is observed. Limited by the sensitivity coefficient of the microcavity to acoustic waves, the MDAP of the proposed sensor is 7.30 mPaHz\textsuperscript{-1/2}, indicating a 26-fold enhancement compared to the transmission method with the same devices. Moreover, the employment of coherent states and heterodyne detection can enhance the response amplitude, providing the possibility of sensitivity up to the sub-micropascal level. With the performance advantage of sensing with both high sensitivity and wide dynamic range, the postselected WGM acoustic sensor offers a promising solution. Potentially, the structure may be applicable to other time-varying signal sensing via WGM microcavities, and broadens various application scenarios, such as magnetic field sensing\cite{M1,M2,hu2024picotesla,li2021cavity}.
	
	\section*{Appendix A. Theoretical description of phase-dracstic and phase-enhanced regions}
	As the application of polarization Mach-Zehnder interferometer, the observable of the postselected WGM acoustic sensor is $\hat{A}=\ket{H}\bra{H}-\ket{V}\bra{V}$, where $\ket{H}$ and $\ket{V}$ are respectively horizontal and vertical polarization state. In polarization modulation step, the probe light is modulated as a linear polarization state with 45\degree, which can be expressed as
	\begin{equation}
		\ket{i}=\frac{1}{\sqrt{2}}(\ket{H}+\ket{V}).
	\end{equation}
	After entering the WGM acoustic sensor, the evolved state of the probe light can be given by
	\begin{equation}
		\ket{\Psi(t)}=B(t)\hat{U}(t)\ket{i},
	\end{equation}
	where $B(t)=\begin{bmatrix}1 & 0 \\ 0 & \sqrt{T(t)}\end{bmatrix}$ denotes the transmission characteristic of the WGM microcavity, and $\hat{U}(t)=\exp[-i\hat{A}\varphi(t)/2]$ is related to phase characteristic $\varphi(t)$. Subsequently, the evolved state is postselected by 
	\begin{equation}	
		\ket{f}=\frac{1}{\sqrt{2}}[\ket{H}-\exp(-i\varepsilon)\ket{V}].
	\end{equation}
	Then the received light intensity can be calculated by $I=|\braket{f|B(t)\hat{U}(t)|i}|^2$, and further expressed as Eq. (\ref{I}).	
	
	When $|\varphi(t)/2|\ll1$, $\hat{U}(t)\approx\hat{I}-i\hat{A}\varphi(t)/2$, where $\hat{I}=\begin{bmatrix}1 & 0 \\ 0 & 1\end{bmatrix}$, and $\braket{f|B(t)\hat{U}(t)|i}\approx \braket{f|B(t)|i}-i\varphi(t)/2\braket{f|B(t)|\hat{A}|i}=\braket{f|B(t)|i}[1-i A_w(t)\varphi(t)/2]$. Here, $A_w$ is the first order scaling factor, i.e.,
	\begin{equation}	
		A_w= \frac{\braket{f|B|\hat{A}|i}}{\braket{f|B|i}},
	\end{equation}
	which is introduced as a scaling factor to the observable by the application of postselection. Under the condition of $|A_w(t) \varphi(t)/2 |\ll1$\cite{dressel2014colloquium,PhysRevA.76.044103}, $\braket{f|B(t)\hat{U}(t)|i}$ can be further expressed as $\braket{f|B(t)|i}\exp[-iA_w(t)\varphi(t)/2]$, which introduce an extraction factor of $\mathrm{Im} A_w$ on the phase characteristic, hence $|\mathrm{Im} A_w|>1$ should be satisfied. Correspondingly, Eq. (\ref{I}) can be approximated to Eq. (\ref{I2}) with $\xi=|\braket{f|B(t)|i}|^2$. 
	
	Based on whether there is an amplification effect, the approach can be divided into phase-drastic and phase-enhanced region. When the two conditions are satisfied, the WGM acoustic sensor is located in phase-enhanced sensing region, and the phase characteristic is amplified by a factor of $\mathrm{Im} A_w \approx \cot(\varepsilon/2)$, which represents the quantitation advantage of postselection\cite{brunner2010measuring,song2023weak}. 
	
	\section*{Appendix B. Performance of acoustic dissipative response}
	The postselected WGM acoustic sensor is also applicable in the detection of dissipative response. Similarly, the response of phase-drastic region can be calculated by
	\begin{equation}
		R_1=\delta_f \frac{\partial (I/I_0)}{M\partial \kappa_{e}}\frac{\partial \kappa_{e}}{\partial s},
	\end{equation}
	where $\partial \kappa_{e}/\partial s$ is the mechanical sensitivity coefficient of the microfiber to acoustic waves, which is also related to the particular coupled microcavity, and the dissipative response of cavity mode with a $Q$ factor of about 10\textsuperscript{8} has an improvement of about one order of magnitude than transmission method, as shown in Fig. \ref{fig5}(a).
	
	Correspondingly in phase-enhanced region, the response can be expressed as 
	\begin{equation}
		\label{r22}
		R_2= \delta_f \frac{\mathrm{Im} A_w}{\xi} \frac{\partial \varphi(t)}{\partial \kappa_{e}}\frac{\partial \kappa_{e}}{\partial s},
	\end{equation}
	and the simulation results in Fig. \ref{fig5}(b) show the improvement could reach 7 orders of magnitude when cavity mode with a $Q$ factor of about 10\textsuperscript{8}. Moreover, the optimal response is stronger than that of transmission method by over 2 orders of magnitude, and the dynamic range can also reach FSR.
	
	\bibliography{ref}	

\end{document}